%% file: sdwcm.tex
\documentstyle[epsf]{europhys}
\input euromacr

\begin{document}
\euro{}{}{}{}
\Date{}
\shorttitle{A. VIROSZTEK \etal FREQUENCY
DEPENDENT CONDUCTIVITY IN SPIN DENSITY WAVES}
\title{Impurity scattering and frequency dependent conductivity\\
in spin density waves}
\author{Attila Virosztek\inst{1,2}, Bal\'azs D\'ora\inst{1}
     \And Kazumi Maki\inst{3}}
\institute{
     \inst{1} Department of Physics, Technical University of Budapest,
              H-1521 Budapest, Hungary\\
     \inst{2} Research Institute for Solid State Physics and Optics
              P.O.Box 49, H-1525 Budapest, Hungary\\
     \inst{3} Department of Physics and Astronomy, University of Southern
              California, Los Angeles, CA 90089-0484, USA}
\rec{}{}
\pacs{
\Pacs{74}{70Kn}{Organic Superconductors}
\Pacs{78}{20$-$e}{Optical properties of bulk materials and thin films}
\Pacs{75}{30Fv}{Spin density waves}
      }
\maketitle
\begin{abstract}
The quasiparticle contribution to the
frequency dependent electric conductivity in the presence of randomly
distributed impurities is calculated
within mean field theory for spin density waves as formed in
quasi one dimensional conductors like
Bechgaard salts. Interchain hopping is taken into account and the
effects of imperfect nesting are considered. In case of an electric field
perpendicular to the chain direction there is no collective contribution
to the conductivity and our results are directly applicable to recent
measurements on both (TMTSF)$_2$PF$_6$ and quenched (TMTSF)$_2$ClO$_4$.
The experimental data are well described by the theory
in its clean limit, with somewhat larger scattering in
the ClO$_4$ salt.
\end{abstract}

Spin density wave (SDW) is one of the ground states of quasi one
dimensional electronic systems, and commonly found at low temperature
in Bechgaard salts (TMTSF)$_2$X with X=PF$_6$, ClO$_4$, etc\cite{IY,Gr}.
Both the appearance of SDW in Bechgaard salts with the pressure dependence
of the SDW transition temperature $T_c$, and the electronic properties
of the SDW are well described in terms of the standard model, where the
approximate nesting of the quasi one dimensional Fermi surface (i.e. the
imperfect nesting), and the repulsive Coulomb interaction between
electrons are the crucial ingredients\cite{Yam}. In particular, the
enhanced gap to $T_c$ ratio compared to the weak coupling BCS value is
interpreted as the consequence of imperfect nesting\cite{HM}. Within this
model the quasi particle energy gap edges depend on the quasi particle
momentum perpendicular to the conducting chains (the ${\bf b}$ direction) as
\begin{equation}
\Delta_{\pm}(p_y)=\pm\Delta+\varepsilon_0\cos(2bp_y),\label{gapedge}
\end{equation}
where $\Delta$ is the order parameter
and $\varepsilon_0$ is the "unnesting" parameter
characterizing the deviation from one dimensionality. Indeed, the above
momentum dependent energy gap is inferred from the magnetoresistance
data\cite{KBM} in the SDW state of (TMTSF)$_2$PF$_6$, and it offers an
explanation of the difference between the energy gap as observed by
transport and optical measurements for a number of density wave
materials\cite{MiVG}. While the optical (pair breaking) gap is still
given by $2\Delta$, dc transport is sensitive to the smaller gap in the
density of states\cite{HM}.

The frequency dependent electric conductivity of the SDW in Bechgaard salts
has been studied over decades\cite{Gr}. Unfortunately however, the
clear understanding of the physical situation still appears to be lacking.
Earlier experiments were mostly done in the geometry ${\bf E}\parallel
{\bf a}$, i.e. the electric field parallel to the chain
direction\cite{sdwpar}. Under these circumstances it is expected that
the phason couples strongly to the electromagnetic field in analogy to
the mean field treatment of CDW by Lee, Rice and Anderson\cite{LRA,VMcdw}.
In SDW, most of the optical weight is shifted to the phason mode and
gives rise to low energy absorption\cite{VMsdw} for $\omega\ll 2\Delta$,
since in the absence of pinning the transverse phason is naturally coupled
to the transverse phonon\cite{VMsound} as well as the photon. However, in
general the phason is pinned by impurities or crystalline defects as
evidenced by the existence of the threshold electric field in the non
Ohmic dc conductivity in (TMTSF)$_2$NO$_3$ and
(TMTSF)$_2$PF$_6$\cite{ST1,ST2}, and therefore the longitudinal phason
gets mixed in as well. On the other hand we have shown\cite{VMsdw} that the
longitudinal phason is almost completely removed from the low frequency
range in an SDW due to the Anderson-Higgs mechanism\cite{PWA}. We believe
that this is the main reason for the lack of a clear optical energy gap in
the data taken in the geometry ${\bf E}\parallel {\bf a}$, since first of
all the optical weight at the gap region $\omega\ge 2\Delta$ is shifted
down to the pinned mode at $\omega\ll 2\Delta$, which in turn is depleted
due to the admixture with the longitudinal phason which is nothing but
the plasmon in a SDW.

In this situation the frequency dependent conductivity perpendicular
to the chains is ideal to measure the optical energy gap, since in the
geometry ${\bf E}\perp {\bf a}$ the electromagnetic field does not
couple to the phason\cite{VMsound}. Within mean field theory we only need
to consider the so called quasiparticle contribution to the conductivity
since no collective contribution is present in this case. We obtain an
expression for the perpendicular electric conductivity which is very
similar to the one derived by Mattis and Bardeen\cite{MB}, and by
Abrikosov {\it et.al.}\cite{AGK} for $s$-wave superconductors except one
crucial point. The authors of both papers\cite{MB,AGK} introduced a
simplification which is valid either in the anomalous skin limit or in
the dirty limit. Unfortunately, for organic conductors and also for
high $T_c$ cuprate superconductors none of the above approximations
applies, since they are usually in the clean limit and the penetration
depth of the ac field is much larger than 10 $\mu$m. An expression for
the SDW conductivity appropriate for the above conditions has already
been obtained in the extreme clean (or collisionless) limit characterized
by infinite quasiparticle mean free path\cite{clean}, but 
for a more complete description the effect of
finite mean free path should certainly be taken into account.

In this report we incorporate the quasiparticle mean free path in terms
of impurity scattering, characterized by a forward ($\Gamma_1$) and
backscattering ($\Gamma_2$) rate\cite{micro}.
These two rates stand for the amplitude of scattering
processes involving electrons
originating from and arriving to the same or different Fermi sheets of the
quasi one dimensional Fermi surface respectively.
Then we compare the theoretical results with
recent experimental data on the SDW of (TMTSF)$_2$PF$_6$ and
(TMTSF)$_2$ClO$_4$\cite{Dressel,Vescoli}. The agreement looks rather
reasonable as long as we use $2\Delta(T=0)\approx 70$cm$^{-1}\approx
100$K for both compound, where $\Delta(T=0)$ is the SDW order parameter
at zero temperature. The order parameter thus determined appears to
be about a factor two larger than the ones determined from transport
measurements, for example $\Delta(T=0)=21$K was deduced in
Ref.\cite{KBM}. This discrepancy however may be understood by
invoking the effect of imperfect nesting\cite{MiVG}
expressed by Eq.(\ref{gapedge}). On the other hand,
the SDW transition temperature $T_c$ can be
reduced from its weak coupling BCS
value of 28K for at least two reasons. First is the effect of imperfect
nesting on $T_c$ which is still within the realm of mean field theory,
and is suggested in particular by the fact that the same energy gap
has been observed in both (TMTSF)$_2$PF$_6$ and (TMTSF)$_2$ClO$_4$.
Second is the suppressing effect of fluctuations, although in our case
they should be at least two dimensional. Indeed, the observed frequency
dependence of the conductivity just above $T_c$ in Bechgaard
salts\cite{Vescoli}
suggests strongly the pseudogap phenomenon.

Our calculation of the frequency dependent conductivity involves
standard diagrammatic treatment of impurity scattering in density
waves\cite{micro}. In order to evaluate the quasiparticle contribution
we dress a single loop of current correlation function by impurity
lines. Self energy corrections are taken into account in the non crossing
approximation, while vertex corrections will vanish due to the constant
scattering rates $\Gamma_1$ and $\Gamma_2$, and to the structure of the
transverse current
\begin{equation}
v_y({\bf p})=\sqrt{2}v_y\sin(bp_y),\label{vy}
\end{equation}
where $v_y=\sqrt{2}bt_b$ is the relevant electron velocity perpendicular
to the chains and $t_b$ is the hopping integral in the ${\bf b}$ direction.
The thermal product corresponding to the current correlation function
is written down, and analytic continuation to the real frequency axis
leads directly to $\sigma_{yy}(\omega)$. For simplicity, we cite here only
the results for perfect nesting, leaving the detailed treatment of
imperfect nesting effects to a forthcoming publication. The frequency
dependent conductivity of a SDW in the ${\bf b}$ direction consists of
two contributions:
\begin{equation}
{\rm Re}[\sigma_{yy}(\omega)]=e^2v_y^2N_0{{\rm Im}[I_n(\omega)+
I_p(\omega)]\over\omega},\label{sigma}
\end{equation}
where $N_0$ is the total density of states (including spin degeneracy)
of the electron system, and the "normal" contribution (involving intraband
processes) is given by
\begin{equation}
{\rm Im}I_n(\omega)=\int_G^\infty {dE\over 2}
\left [ \tanh{E+\omega\over 2T}-\tanh{E\over 2T}\right ]
{\rm Re}\{ F[u(E+\omega),u(E)]-F[u(E+\omega),u^*(E)]\},\label{In}
\end{equation}
while the so called pair breaking (interband) contribution is present only
for frequencies larger than the gap $2G$ in the quasiparticle spectrum:
\begin{equation}
{\rm Im}I_p(\omega > 2G)=\int_G^{\omega-G}{dE\over 2}\tanh{\omega-E\over 2T}
{\rm Re}\{ F[u(\omega-E),-u^*(E)]-F[u(\omega-E),-u(E)]\}.\label{Ip}
\end{equation}
In the above two expressions the function $F$ of two variables is given by
\begin{equation}
F(u,u^\prime)={-\left [ 1-{1-uu^\prime\over\sqrt{1-u^2}
\sqrt{1-(u^\prime)^2}}\right ]\over\Delta [\sqrt{1-u^2}+
\sqrt{1-(u^\prime)^2}]-\Gamma_1},\label{Fy}
\end{equation}
and the function $u(E)$ is defined by
\begin{equation}
{E\over\Delta}=u\left ( 1-{\alpha\over\sqrt{1-u^2}}\right ).\label{u}
\end{equation}
Here the pair breaking parameter $\alpha=\Gamma/\Delta$ with $\Gamma=
\Gamma_1+{1\over 2}\Gamma_2$, while the gap is related to the order
parameter as $G=\Delta u_0^3$ where $u_0^2=1-\alpha^{2/3}$, unless we
are in the gapless regime very close to $T_c$\cite{micro}.

We note here, that the calculation of the quasiparticle term for the
conductivity in the chain direction proceeds rather similarly, except
for the following two differences: in order to obtain Re$[\sigma_{xx}
(\omega)]$ we first need to replace $v_y$ by the Fermi velocity $v_F$
in Eq.(\ref{sigma}), and second, because of the important vertex corrections
in this case we need to use the following $F$ function
\begin{equation}
F_x(u,u^\prime)={-\left [ 1-{1-uu^\prime\over\sqrt{1-u^2}
\sqrt{1-(u^\prime)^2}}\right ]\over\Delta [\sqrt{1-u^2}+
\sqrt{1-(u^\prime)^2}]-(\Gamma_1-{1\over 2}\Gamma_2)\left [1-
{1-uu^\prime\over\sqrt{1-u^2}\sqrt{1-(u^\prime)^2}}\right ]}\label{Fx}
\end{equation}
in place of Eq.(\ref{Fy}).

\begin{figure}
\epsfxsize=14cm
\epsfysize=10cm
\epsffile{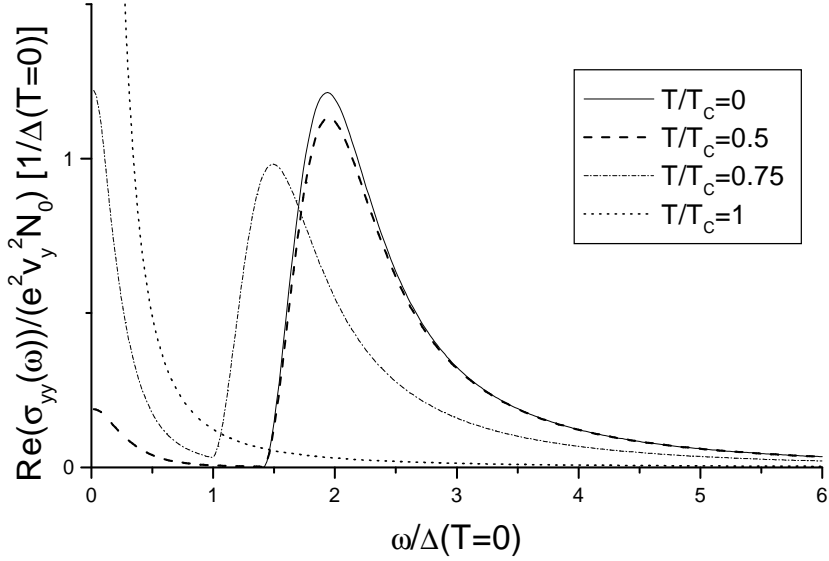}
\caption{Real part of the normalized frequency dependent conductivity
in the direction perpendicular to the chains for several temperatures
relative to the critical temperature of the spin-density wave
transition $T_c$. The electron forward- and backscattering rates due
to impurities are $\Gamma_1=2\Gamma_2=0.08\Delta(T=0)$, corresponding to
a pair breaking parameter $\alpha=0.1$ at zero temperature.}
\label{fig1}
\end{figure}

Limiting ourselves to the case ${\bf E}\perp{\bf a}$ (i.e. we use
Eq.(\ref{Fy}) for the $F$ function), we have evaluated the
real part of the
conductivity as a function of frequency as given by Eq.(\ref{sigma})
for $T/T_c=0$, 0.5, 0.75 and 1, for two sets of values of the impurity
scattering rates leading to $\alpha=0.1$ and 0.25 at zero temperature.
The results are shown in Fig.1 and Fig.2 respectively. It is demonstrated
that the normal state ($T/T_c=1$) Drude peak given by Eq.(\ref{sigma})
evaluated for $\Delta=0$:
\begin{equation}
{\rm Re}[\sigma_{yy}(\omega)]=e^2v_y^2N_0{\Gamma_1+\Gamma_2\over
\omega^2+(\Gamma_1+\Gamma_2)^2},\label{sigman}
\end{equation}
is gradually frozen out as the temperature is lowered, transferring
all the spectral weight to the pair breaking peak around $2\Delta$
at zero temperature. The values of the impurity scattering rates were
chosen so that our figures would be representative of the clean limit
situation, and would most closely resemble the width and
shape of the experimental data\cite{Vescoli} for (TMTSF)$_2$PF$_6$ and
(TMTSF)$_2$ClO$_4$. Unfortunately the detailed quantitative comparison
with experiments is hindered by contributions to the frequency dependent
conductivity not related to the SDW discussed here. However, we can
estimate for example the ratio of the peak width (at half maximum) to
the peak position (which is 0.45 on Fig.1) at low temperature, and obtain
0.45 for the PF$_6$ and 0.50 for the ClO$_4$ salt.
Our results confirm that these
Bechgaard salts are indeed in the relatively clean limit. We note here
in passing that if the current flows in the chain direction (${\bf E}
\parallel{\bf a}$), than the transport lifetime $\Gamma_1+\Gamma_2$ in
Eq.(\ref{sigman}) gets replaced by $2\Gamma_2$ due to the vertex
corrections, since only backscattering can cause current damping in the
chain direction.

\begin{figure}
\epsfxsize=14cm
\epsfysize=10cm
\epsffile{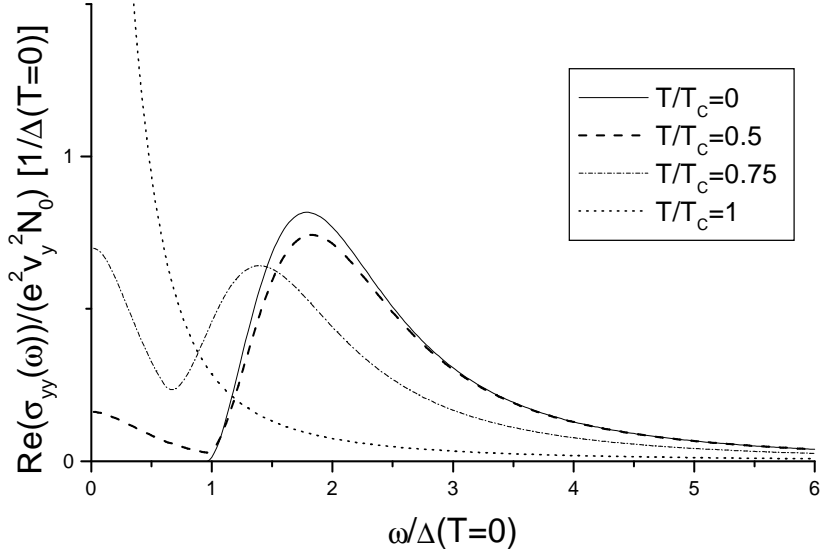}
\caption{Same as in Fig.1, except that here $\Gamma_1=2\Gamma_2=0.2\Delta
(T=0)$, corresponding to a zero temperature value of $\alpha=0.25$.}
\label{fig2}
\end{figure}

Perhaps some additional analytic results derivable from the previous
equations in the experimentally relevant clean limit and at low
temperatures are of interest.
First we can evaluate the dc conductivity as
\begin{equation}
\sigma_{ii}^{dc}=e^2v_i^2N_0{2T\over 3\Delta\Gamma u_0}e^{-G/T},\label{dc}
\end{equation}
where $i=x,y$ and $v_x=v_F$. It is worth to note that the ratio of the
dc resistivities in the two directions $v_y^2\rho_{yy}/v_F^2\rho_{xx}$
is unity for zero temperature, while takes on the value ${1\over 2}
(1+\Gamma_1/\Gamma_2)$ for $T\rightarrow T_c$, indicating a temperature
dependence of this ratio even though the scattering rates are themselves
temperature independent. Also the shape of the central peak which
freezes out for $T\rightarrow 0$ turns out to
be different from the usual Lorentzian. If the scattering is not too
small, the conductivity exhibits an unconventional $\omega^{-1/2}$
behavior in the frequency range $T\ll\omega\ll G$:
\begin{equation}
{\rm Re}[\sigma_{ii}(\omega)]=\sigma_{ii}^{dc}\sqrt{\pi T\over 4\omega}.
\label{sqrt}
\end{equation}
Finally, we can evaluate the pair breaking contribution near the
threshold $\omega\simeq 2G$:
\begin{equation}
{{\rm Re}[\sigma_{yy}(\omega)]\over e^2v_y^2N_0}={8\over 9\alpha^{2/3}}
{(\omega-2G)^2\over\Gamma G\omega},\label{thy}
\end{equation}
and
\begin{equation}
{{\rm Re}[\sigma_{xx}(\omega)]\over e^2v_F^2N_0}={1+5\Gamma_2/\Gamma_1+
(3\Gamma_2/2\Gamma_1)^2\over 36\alpha^{2/3}}
{(\omega-2G)^2\over\Gamma_1 G\omega}.\label{thx}
\end{equation}

In summary, we have obtained the expression for the frequency dependent
conductivity in a SDW for ${\bf E}\perp{\bf a}$ within mean field
theory. The effect of the quasiparticle lifetime is incorporated in
terms of impurity scattering. The present results are shown on Figs. 1
and 2 in the weak scattering limit, and appear to describe
the observed conductivity of both (TMTSF)$_2$PF$_6$ and
(TMTSF)$_2$ClO$_4$ reasonably well.
It is highly desirable to test how the Drude-like
peak disappears at low temperatures. Also the present model highlights
the different transport lifetimes for ${\bf E}\parallel {\bf a}$ and
${\bf E}\perp{\bf a}$, which is indeed consistent with some of the
magnetotransport measurements\cite{KBM}.

\stars

This work was supported by the Hungarian National Research Fund under
grant numbers OTKA T020030 and T015552,
and by the Ministry of
Education under grant number FKFP 0029/1999. One of us (A. V.)
acknowledges greatfully the hospitality during his stay at the University
of Southern California.
\end{document}

%% file: euromacr.tex
\def\etal{{\hbox{{\tenit\ et al.\/}\tenrm :\ }}}

\def\And{{\rm and\ }}

\def\stars{\bigskip\centerline{***}\medskip}

\newif\ifboo \boofalse

\def\Review#1{\boofalse{\it #1},}
\def\Name#1{{\sc #1},}
\def\Vol#1{\ifboo Vol. {\bf #1}\else{\bf #1}\fi}
\def\Year#1{\ifboo #1\else(#1)\fi}
\def\Book#1{\bootrue{\it #1},}
\def\Page#1{\ifboo {\rm p. #1}\else{\rm #1}\fi}